\documentclass[preprint,pre,amssymb,amsmath,showpacs,floatfix]{revtex4}
\usepackage{graphicx}
\usepackage{amssymb}
\usepackage{amsmath}
\usepackage{subfigure}

\begin{document}

\title{Fastest Learning in Small-World Neural Networks }
\author{D.~Simard, L.~Nadeau, and 
H.~Kr\"{o}ger$\footnote{Corresponding author, Email: hkroger@phy.ulaval.ca}$ }
\affiliation{ {\small\sl D\'{e}partement de Physique, Universit\'{e}
Laval, Qu\'{e}bec, Qu\'{e}bec G1K 7P4, Canada} }
\begin{abstract}
We investigate supervised learning in neural networks. We consider a multi-layered
feed-forward network with back propagation. We find that the network
of small-world connectivity reduces the learning error and learning
time when compared to the networks of regular or random connectivity. 
Our study has potential applications in the domain of data-mining, image processing, 
speech recognition, and pattern recognition.
\end{abstract}

\maketitle

\section{Introduction}
The concept of small-world networks and the variant of
scale free networks has become very popular recently
\cite{Watts98,Albert99,Barabasi99,Huberman99,Albert00,Jeong00,Kleinberg00,Strogatz01,Watts02}, 
after the discovery that such networks are realized in diverse areas as the 
organization of human society (Milgram's experiment)
\cite{Watts98}, in the WWW \cite{Albert99,Huberman99}, in the
internet \cite{Faloutsos99}, in the distribution of electrical power
(western US) \cite{Watts98}, and in the metabolic network of the
bacterium {\it Escherichia coli} \cite{Strogatz01,Wagner01}.

According to Watts and Strogatz, a small-world network is
characterized by a clustering coefficient $C$ and a path
length $L$ \cite{Watts98}. The clustering coefficient measures the
probability that given node $a$ is connected to nodes $b$ and $c$ then also 
nodes $b$ and $c$ are connected. The shortest path
length from node $a$ to node $b$ is the minimal number of
connections needed to get from $a$ to $b$. A small-world
network is defined by two properties. First, the 
average clustering coefficient $C$ is larger 
than for a corresponding random network with the same number of 
connections and nodes. The clustering coefficient expected for a regular rectangular
grid network is zero while for a random network the probability $p$ of 
connection of two nodes is the same for neighboring nodes as for distant nodes. 
Second, the average path length $L$
scales like $\log N$, where $N$ is the number of nodes. For regular (grid)
networks $L$ scales as $N^d$ where $d$ is the dimension of the space and 
for random networks $L$ scales as $\log N$. As result, 
$C$ is large and $L$ is small in a small-world network. 

\begin{figure*}[ht]
\vspace{0.0cm}
\begin{center}
\includegraphics[scale=0.6,angle=-90]{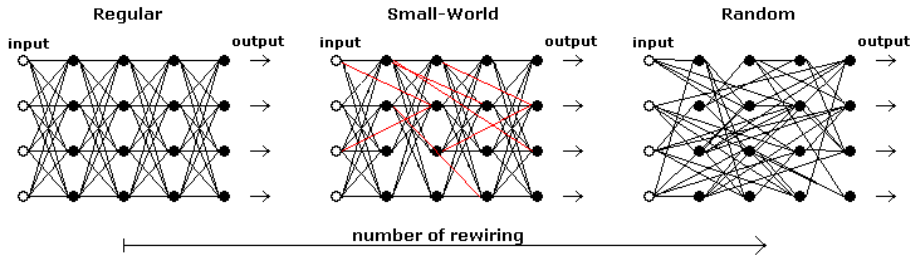}
\vspace{0.0cm}
\end{center}
\caption{Scheme of network connection topologies obtained by randomly 
cutting and rewiring connections, starting from regular net (left) 
and going to random net (right). The small-world network is located 
somewhere in between.} 
\label{fig:connectivity}
\end{figure*}

In studies of small-world neural networks it has turned out that this architecture 
has potentially many advantages. In a Hodgkin-Huxley network the small-world
architecture has been found to give a fast and synchronized response
to stimuli in the brain \cite{Corbacho00}. In associative memory
models it was observed that the small-world architecture yields the
same memory retrieval performance as randomly connected networks,
using only a fraction of total connection length \cite{Bohland01}.
Likewise, in the Hopfield associative memory model the small-world
architecture turned out to be optimal in terms of memory storage
abilities \cite{Labiouse02}. With a integrate-and-fire type of
neuron it has been shown \cite{Sola04} that short-cuts permit
self-sustained activity. 
A model of neurons connected in small-world topology has been used to explain 
short bursts and seizure-like electrical activity in epilepsy \cite{Netoff04}.
In biological context some
experimental works seem to have found scale-free and small-world
topology in living animals like in the macaque visual cortex\cite{Sporns00},
in the cat cortex \cite{Sporns00} and even in networks 
of correlated human brain activity measured 
via functional magnetic resonance imaging \cite{Dante04}. 
The network of cortical neurons in the brain has sparse long ranged connectivity, 
which may offer some advantages of small world connectivity \cite{Laughlin03}.

\begin{figure*}[ht]
\vspace{0.0cm}
\begin{center}
\includegraphics[scale=0.55,angle=90]{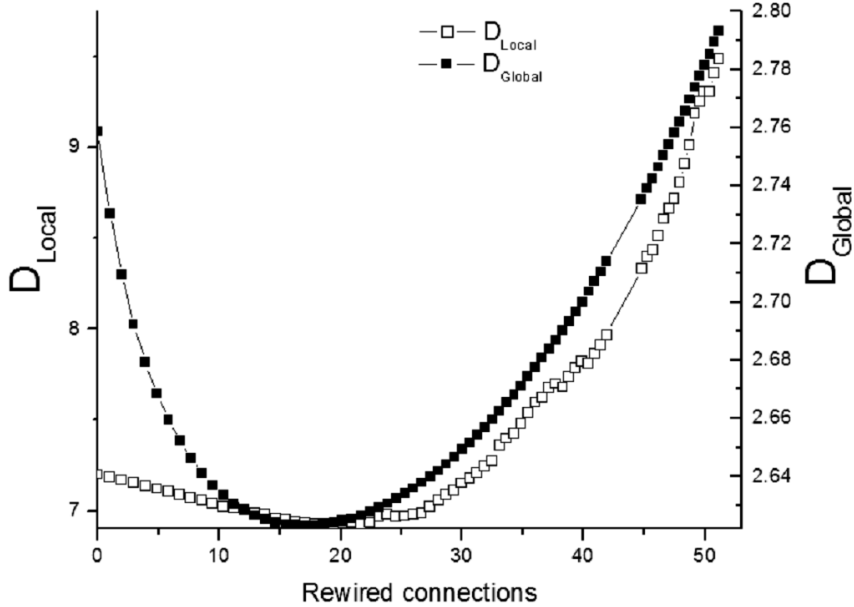}
\vspace{0.0cm}
\end{center}
\caption{$D_{local}$ and $D_{global}$ versus
number of rewirings. Networks with 5 neurons per layer and 5 layers. }
\label{fig:DLocGlob}
\end{figure*}

\section{Model and geometry of network}
In the present work we study supervised learning with
back-propagation in a multi-layered feed-forward network. Around
1960, the Perceptron model \cite{Rosenblatt62} has been widely
investigated. The information (representing action potentials
propagating in axons and dendrites of biological neurons) feeds
in forward direction. The Perceptron
model has been extended to include several layers. 
E.g., a convolutional network, consisting of seven layers plus one input layer 
has been used for reading bank cheques \cite{LeCun98}.  
The task of learning consists of finding optimal weights $w_{ij}$ between
neurons (representing synaptic connections) such that for a given
set of input and output patterns (training patterns) the network
generates output patterns as close as possible to target
patterns. We used the algorithm of back propagation
\cite{Rummelhart86} to determine those weights. There are alternative, potentially faster
methods to determine those weights like, e.g. simulated annealing. 
Here we aim to compare different network architectures with respect
to learning, using as reference a standard algorithm to determine
those weights.

\begin{table}[hb]
\centering
\begin{tabular}{|c|c|c|c|c|c|c|c|c|}
\hline
$N_{neuron}$ & $N_{layer}$ & $N_{nodes}$ & $K_{conn}$ & $N^{min}_{rewire}$ &
 $D^{min}_{global}$ & R & $\Sigma$ & S \\
\hline
 5 & 5 & 25 & 100 & 19 $\pm$ 5 & 2.61 $\pm$ .02 & 0.19 $\pm$ .05 & 0.38 $\pm$ .02 & 0.81 $\pm$ .01 \\
\hline
 5 & 8 & 40 & 175 & 28 $\pm$ 5 & 3.25 $\pm$ .02 & 0.17 $\pm$ .02 & 0.34 $\pm$ .06 & 0.88 $\pm$ .01 \\
\hline 
 10 & 10 & 100 & 900 & 400 $\pm$ 100 & 3.40 $\pm$ .01 & 0.44 $\pm$ .11 & 0.30 $\pm$ .02 & 0.74 $\pm$ .002 \\
\hline 
 15 & 8 & 120 & 1575 & 830 $\pm$ 50 & 3.20 $\pm$ .01 & 0.53 $\pm$ .03 & 0.29 $\pm$ .003 & 0.67 $\pm$ .002\\
\hline 
 15 & 15 & 225 & 3150 & 2300 $\pm$ 300 & 3.75 $\pm$ .01 & 0.73 $\pm$ .09  & 0.29 $\pm$ .01 & 0.69 $\pm$ .002 \\
\hline 
\hline
\end{tabular}
\caption{Network parameters. $N^{min}_{rewire}$ and $D^{min}_{global}$ correspond to position where 
both $D_{local}$ and $D_{global}$ are small. Scaling of $R$ (Eq.(\ref{eq:ScalingR})) 
and of $D_{global}$ (Eq.(\ref{eq:ScalingDglobal})). } 
\label{tab:Scaling}
\end{table}

\begin{figure*}[ht]
\vspace{0.0cm}
\begin{center}
\includegraphics[scale=0.5,angle=-90]{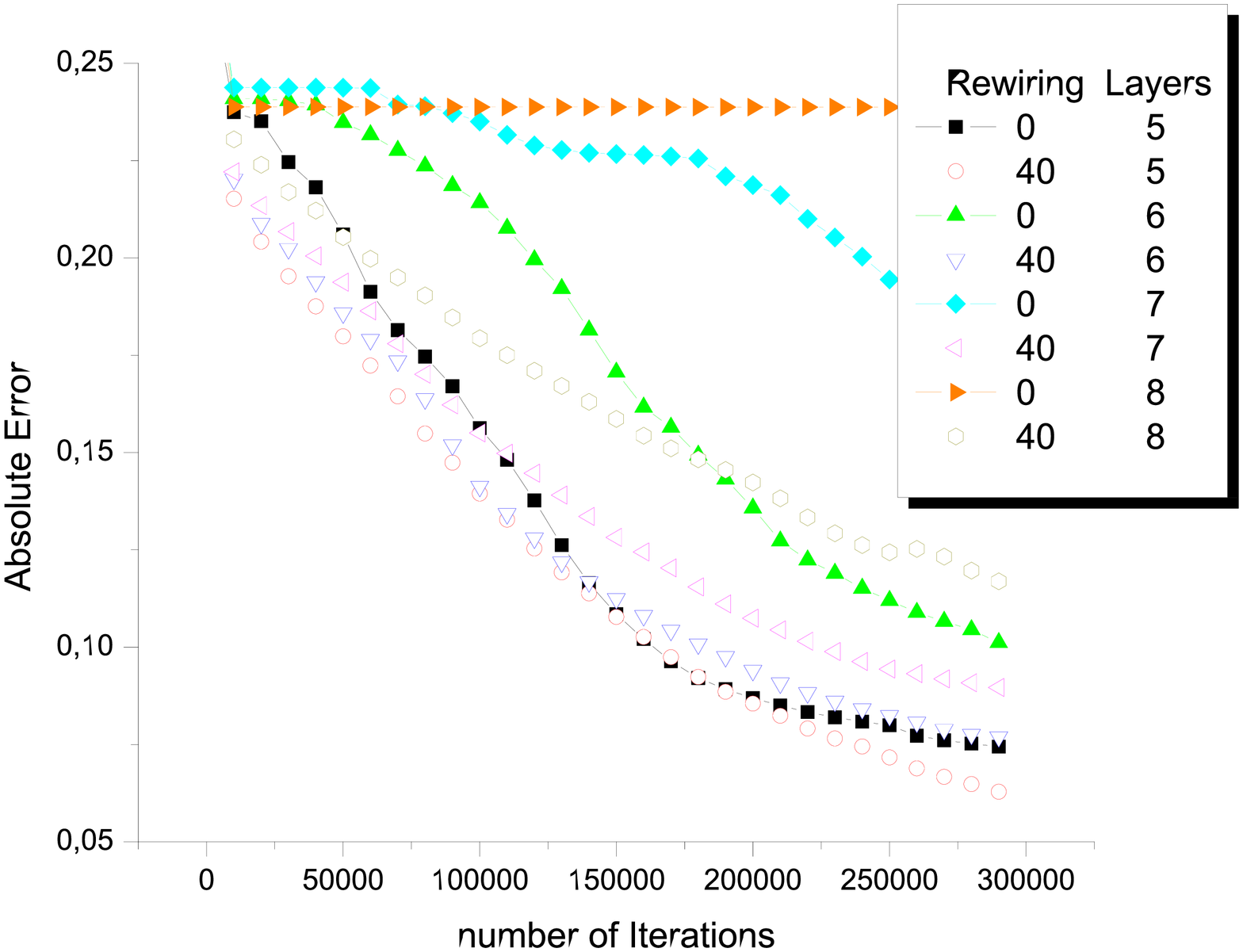}
\vspace{0cm}
\end{center}
\caption{Network of 5 neurons per layer and different number of layers.
Learning of 20 patterns, learning rate 0.02 (back-propagation), training by 10 statistical tests. } 
\label{fig:Function_Layer}
\end{figure*}

We change the architecture of neural connections from regular to random 
architecture, while keeping the number $K_{conn}$ of connections fixed. Initially, 
neurons are connected feed-forward, i.e. each neuron of a given layer connects to all
neurons of the subsequent layer. We make a random draw of two nodes which are 
connected to each other. We cut that "old" link. In order to create a "new" link, 
we make another random draw of two nodes. If those nodes are already connected 
to each other, we make further draws until we find two unconnected nodes. 
Then we create a "new" link between those nodes.
Thus we conserve the total number $K_{conn}$ of links. The number of links in 
a regular network is given by $K_{conn} = N_{neuron}^{2} \times (N_{layer}-1)$, where $N_{neuron}$
denotes the number of neurons per layer and $N_{layer}$ denotes the number layers. 
In this way we create some connections between nodes in distant layers, 
i.e. short-cuts and  the topology changes gradually (see
Fig.[\ref{fig:connectivity}]). In particular, while the initial connectivity 
is regular, after many rewiring steps the topology becomes
random. Somewhere in between lies the small-world architecture
\cite{Watts98}. We consider networks consisting of one input layer, 
one output layer and one or more hidden layers. While the standard Perceptron 
model has only two layers, here we explore the use of more hidden layers, because 
only then the path length $L$ can become small and the network "small-world".  

Instead of measuring the network architecture by the
functions $C$ and $L$, we have used the functions of local and
global connectivity length $D_{local}$ and $D_{global}$, respectively, 
introduced by Marchiori et al. \cite{Latora,Marchiori00}. Those functions are more 
suitable because they allow to take into account the connectivity matrix and 
the weight matrix in networks and also to treat networks with some unconnected 
neurons (often nodes are unconnected due to rewiring). 
$D_{local}$ and $D_{global}$ are defined via the concept of global and local 
efficiency \cite{Latora,Marchiori00}. For a graph $G$ its efficiency is 
defined by
\begin{equation}
E_{global}(G) = \frac{1}{N(N-1)} \sum_{i \neq j \in G} \epsilon_{ij} ~ , 
~~~ \epsilon_{ij}=1/d_{ij} ~ , 
\end{equation}
where $d_{ij}$ denotes the shortest path length between vertices $i$ 
and $j$ and $N$ is the total number of vertices. Moreover, one defines 
the local efficiency as average efficiency of subgraphs $G_{i}$ 
(the subgraph $G_{i}$ is usually formed by the neurons directly connected 
to neuron $i$. We have included also all neurons occuring in the same 
layer as neuron $i$).
\begin{equation}
E_{local} = \frac{1}{N} \sum_{i \in G} E(G_{i}) ~ .
\end{equation}
The connectivity length is defined by the inverse of efficiency,
\begin{equation}
D_{global} = 1/E_{global} ~ , ~~~ D_{local} = 1/E_{local} ~ .
\end{equation}
It has been shown that $D_{global}$ is similar to $L$, and $D_{local}$
is similar to $1/C$ \cite{Marchiori00}, although this is not an exact relation. 
Thus, if a network is small-world, both, $D_{local}$ and $D_{global}$ should become
small. Fig.[\ref{fig:DLocGlob}] shows $D_{local}$ and $D_{global}$ as a function of 
the number of rewiring steps for a network of 5 neurons per layer and 5 layers. 
One observes that both, $D_{local}$ and $D_{global}$, become small at 
about $N_{rewire} \approx 20$. The position of minima for other networks are shown in 
Tab.[\ref{tab:Scaling}].

Small world networks are characterized by the scaling law $L \propto \log(N_{nodes})$. 
From the similarity of $D_{global}$ and $L$, we expect  
\begin{equation}
D_{global} \propto ~ \log(N_{nodes}), ~~~ (\mbox{for} ~ N_{nodes} \to \infty) ~ .
\label{eq:ScalingDglobal}
\end{equation}
In Tab.[\ref{tab:Scaling}] we display $S=D^{min}_{global}/\log(N_{nodes})$, 
where $D^{min}_{global}$ denotes the value of $D_{global}$ at the position 
where both $D_{local}$ and $D_{global}$ are small.
The onset of scaling is observed when networks become larger. 
From our data we observed another scaling law for the variable 
$R= N^{min}_{rewire}/K_{conn}$. Our data are consistent with
\begin{equation}
R = R_{0} + \Sigma ~ \log(K_{conn}) ~~~ (\mbox{for} ~ N_{nodes} \to \infty) ~ ,
\label{eq:ScalingR}
\end{equation}
where $R_{0}$ and $\Sigma$ are constants. 
In Tab.[\ref{tab:Scaling}] we display $\Sigma$. This law says that the number of 
rewirings associated with the minimum of $D_{local}$ and $D_{global}$, i.e. 
small world connectivity, measured in terms of the number of connections of 
the regular network, increases with the number of connections in a logarithmic way. 

\begin{figure*}[ht]
\vspace{0.0cm}
\begin{center}
\includegraphics[scale=0.9]{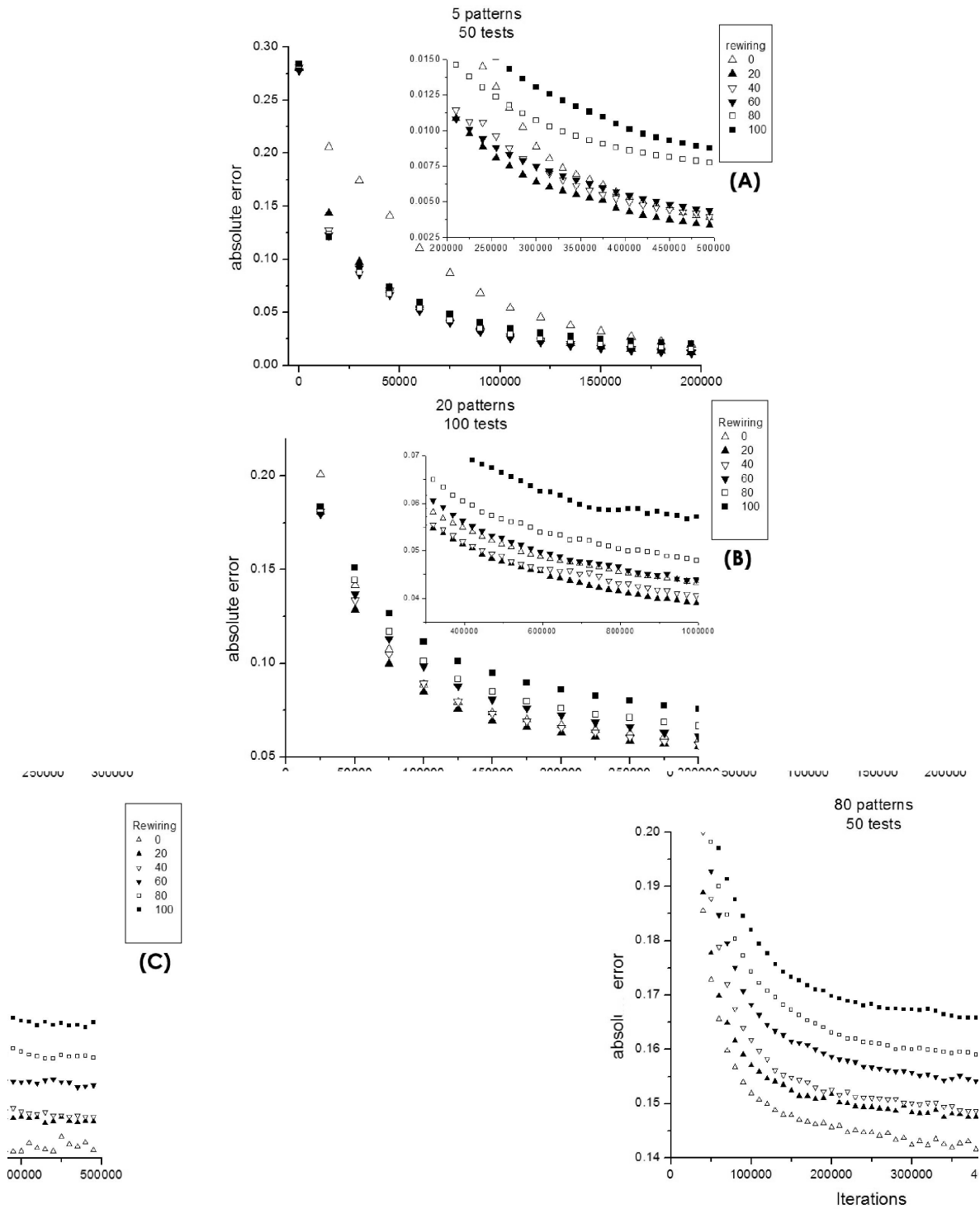}
\vspace{0.0cm}
\end{center}
\caption{Network of 5 neurons per layer and 5 layers. 
Learning of 5 patterns (a), 20 patterns (b) and 80 patterns (c).} 
\label{fig:Function_Pattern}
\end{figure*}

\section{Learning} 

We trained the networks with random binary input and output patterns. 
Our neurons are sigmoid real-valued units.
The learning time, defined as the number of iterations it takes until the error
of training becomes smaller than a given error tolerance,  
depends on the error tolerance. In the following we will present the learning error instead of 
learning time.
\begin{figure*}[ht]
\vspace{0.0cm}
\begin{center}
\includegraphics[scale=0.5,angle=90]{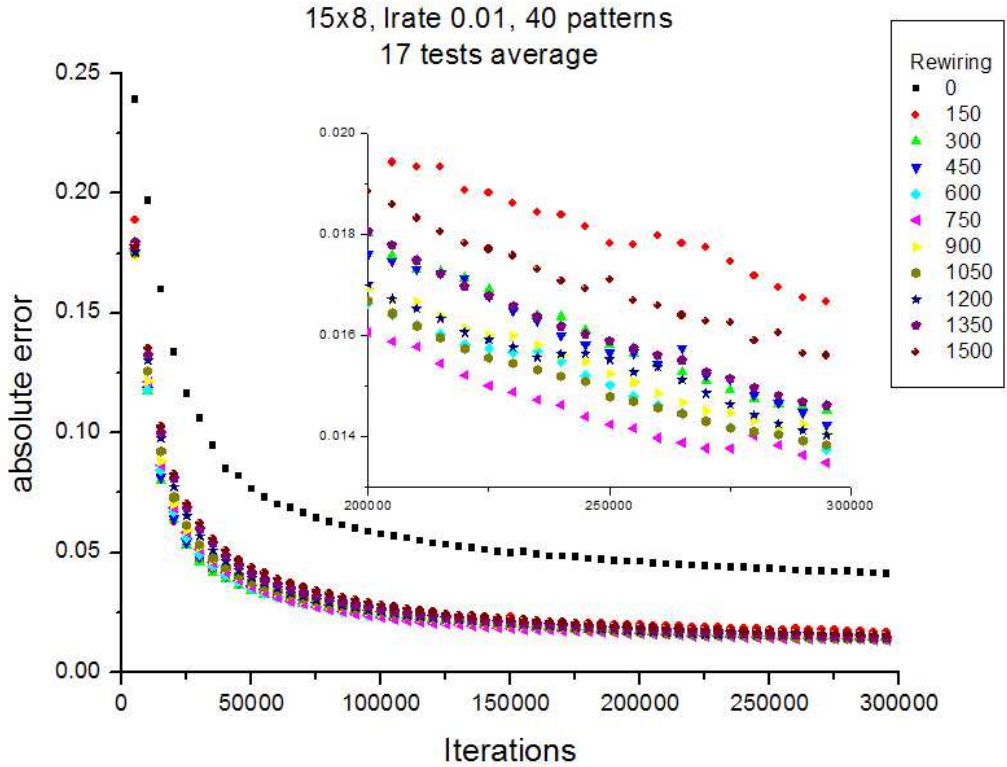}
\vspace{-0.5cm}
\end{center}
\caption{
Network of 15 neurons per layer with 8 layers. 
Network was trained with 40 patterns by 17 statistical tests.
}
\label{fig:Function_Neurons}
\end{figure*}
The effect on learning due to a variation of the number of layers and the 
effect of rewiring, i.e. introducing short-cuts, is shown in 
Fig.[\ref{fig:Function_Layer}]. It shows the absolute error, which 
is an average of the absolute error over the neurons, patterns
and statistical tests. We have considered a network of 5 neurons per layer 
and varied the number of layers from 5 to 8. 
We compare the regular network ($N_{rewire}=0$) 
with the network of $N_{rewire}=40$. We find that for up to a few thousand 
iterations the network with $N_{rewire}=40$ gives a smaller error compared to the 
regular network. In the regime of many iterations, one observes that the 
5-layer regular network learns (i.e. the error decreases with iterations), 
while the regular 8-layer network does not learn at all. In contrast, after 
40 rewirings all networks do learn. The improvement over the regular network 
is small for 5 layers, but major for 8 layers. This indicates that the  
learning improvement due to small-world architecture becomes more efficient 
in the presence of more layers. In the case of 8 layers, 
the error curve for $N_{rewire}=40$ is in the regime where 
$D_{local}$ and $D_{global}$ are both small (see Tab.[\ref{tab:Scaling}]), 
i.e. close to the small-world architecture.
 
\begin{figure*}[ht]
\vspace{0.0cm}
\begin{center}
\includegraphics[scale=0.7]{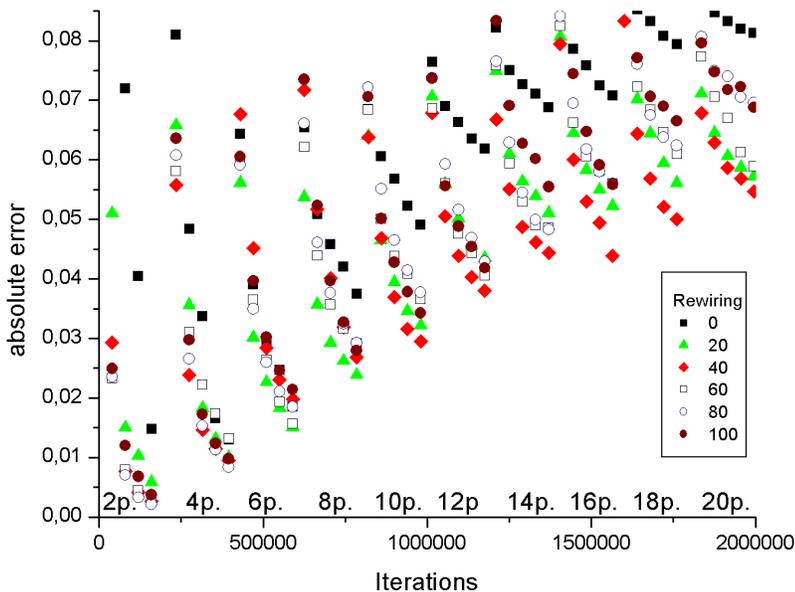}
\vspace{0cm}
\end{center}
\caption{Network of 5 neurons per layer and 5 layers.  
Learning by adding patterns sequentially in groups. 
Network was trained by 20 statistical tests.} 
\label{fig:AddPattSeq}
\end{figure*}

The effect on learning when changing the number of learning patterns and the 
rewiring of the network is shown in Fig.[\ref{fig:Function_Pattern}] 
for a network of 5 neurons per layer and 5 layers. When learning 5 patterns 
(Fig.[\ref{fig:Function_Pattern}a]) in the domain of few iterations (1000-5000), 
rewiring brings about a substantial reduction in the error compared to the 
regular network ($N_{rewire}=0$) and also to the random network ($N_{rewire}>100$). 
For very many iterations (about 500000) there is little 
improvement with rewirings compared to the regular architecture. For learning 
of 20 patterns (Fig.[\ref{fig:Function_Pattern}b]) the behavior is similar, but the 
reduction of error in the presence of 20 rewirings is more substantial 
for many iterations. When learning 80 patterns Fig.[\ref{fig:Function_Pattern}c] 
we see that the error is large, we are near or beyond the storage capacity. 
In this case the regular architecture is optimal and rewiring brings no advantage.

\begin{figure*}[ht]
\vspace{0.0cm}
\includegraphics[scale=0.50,angle=-90]{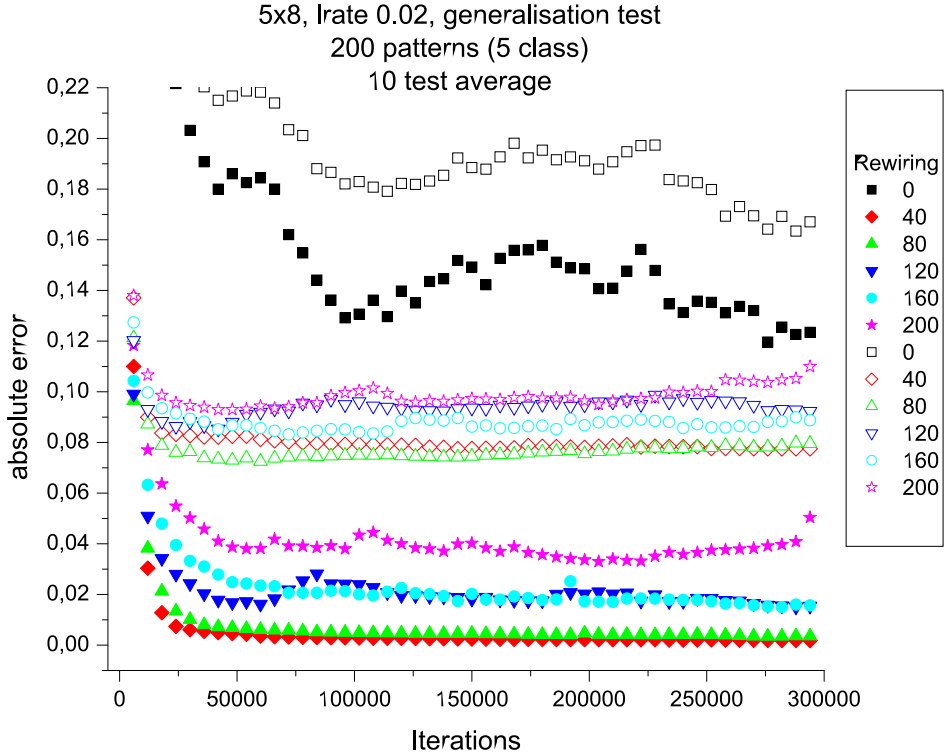}
\vspace{0.0cm} 
\caption{Generalization. Network of 5 neurons per layer and 8 layers. 
Full symbols correspond to trained patterns, empty symbols correspond 
to untrained patterns.} 
\label{fig:Gen5x8}
\end{figure*}

The influence of the number of neurons per layer on learning is depicted in 
Fig.[\ref{fig:Function_Neurons}]. We compare (a) a network of 5 neurons per layer and 8 layers 
with a network (b) of 15 neurons per layer and 8 layers. We trained the network with 40 patterns 
in both cases. In case (a) (not shown) we found that the error as function 
of rewiring has a minimum at $N_{rewire} \approx 30$ which is in coincidence with the minimum of 
$D_{local}$ and $D_{global}$ given in Tab.[\ref{tab:Scaling}], i.e in the small-world regime.
This gives a clear improvement over the regular architecture ($N_{rewire}=0$) 
and the random architecture ($N_{rewire}>100$). 
In case (b) the learning error is shown in Fig.[\ref{fig:Function_Neurons}]. 
One observes that the learning error has a minimum at 
$N_{rewire} \approx 750$, which is close to the minimum of $D_{local}$ and 
$D_{global}$ at $N_{rewire} \approx 830$ (see Tab.[\ref{tab:Scaling}]), 
also near the small-world regime.  
  
So far we studied learning when the training patterns were given all together 
in the beginning. However, sometimes one is interested in changing strategies or parameters 
during training (enforced learning). In order to see if the small world architecture 
gives an advantage in such scenarios, we studied learning by adding patterns 
sequentially in groups of 2 patterns. 
For a network of 5 neurons per layer and 5 layers, Fig.[\ref{fig:AddPattSeq}]   
shows the learning error versus the number of rewirings.
Also here one observes a distinctive gain by introducing some rewirings 
(optimum at $N_{rewire} \approx 40$) over the regular network ($N_{rewire}=0$) 
and the random network ($N_{rewire}>100$).

In the last experiment we studied generalization in a network of 5 neurons per layer 
and 8 layers. We trained the network to put patterns into classes. We considered 5 classes: 
A pattern belongs to class 1, if neuron 1 in the input layer has the highest value of all 
neurons in this layer. Class 2 corresponds to neuron 2 having the highest value. etc. 
The classification of 200 patterns achieved by the network as function of connectivity is shown 
in Fig.[\ref{fig:Gen5x8}]. It turns out the network with some rewirings gives improvement 
over the regular and also random network architecture. We observe that the generalization 
error has a minimum at  $N_{rewire} \approx 40$, which is in the regime where 
$D_{local}$ and $D_{global}$ are small (see Tab.[\ref{tab:Scaling}]), 
i.e. in the regime of small-world architecture.

In summary, we observed that the network with some rewirings., i.e. short-cuts, 
gives reduced learning errors.
The minimum of the error as function of rewiring lies in the neighborhood of the minimum of 
$D_{local}$ and $D_{local}$ which indicates small-world architecture. However, 
while $D_{local}$ and $D_{global}$ depend only on the connectivity of the network, 
learning error and learning time depend on other parameters like the number of learning 
patterns, or the mode of learning, e.g. standard learning or enforced learning.
We believe that our results have important implications on artificial
intelligence and artificial neural networks. Neural networks are
being widely used as computational devices for optimization problems
based on learning in applications like pattern recognition, image processing, 
speech recognition, error detection, and quality control in industrial production, like of car
parts or sales of air-line tickets. Our study reveals a clear
advantage and suggests to use small-world networks in such
applications. Our study may have potential applications in the domain 
of data-mining.

\vspace{0.5cm}

\noindent {\bf Acknowledgements} \\
This work has been supported by NSERC Canada. H.K. is grateful for
discussions with Y. DeKoninck, A. Destexhe and M. Steriade.


\begin{thebibliography}{9}

\bibitem{Watts98} Watts, D.J. \& Strogatz, S.H.
{\it Nature} {\bf 393}, 440-442 (1998).

\bibitem{Albert99} Albert, R., Jeong, H. \& Barab\'{a}si, A.L.
{\it Nature} {\bf 401}, 130-131 (1999).

\bibitem{Barabasi99} Barab\'{a}si, A.L. \& Albert, R.
{\it Science} {\bf 286}, 509 (1999).

\bibitem{Albert00} Albert, R., Jeong, H. \& Barab\'{a}si, A.L.
{\it Nature} {\bf 406}, 378-382 (2000).

\bibitem{Jeong00} Jeong, H., Tombor, B., Albert, R., Oltvai, Z.N. \& Barab\'{a}si, A.L.
{\it Nature} {\bf 407}, 651-654 (2000).

\bibitem{Huberman99} Huberman, B.A. \& Adamic, L.A.
{\it Nature} {\bf 401}, 131 (1999).

\bibitem{Kleinberg00}  Kleinberg, J.M.
{\it Nature} {\bf 406}, 845 (2000).

\bibitem{Strogatz01} Strogatz, S.H.
{\it Nature} {\bf 410}, 268-276 (2001).

\bibitem{Watts02} Watts, D.J., Dodds P.S. \& Newman, M.E.J.
{\it Science} {\bf 296}, 1302 (2002).

\bibitem{Faloutsos99} Faloutsos, M., Faloutsos, P. \& Faloutsos, C.
{\it Comp. Comm. Rev.} {\bf 29}, 251-262 (1999).

\bibitem{Wagner01} Wagner, A.
{\it Proc. R. Soc. London} {\bf B268}, 1803-1810 (2001).

\bibitem{Corbacho00} Corbacho, H.R., Lago-Fern\'{a}ndez, 
F., Sig\"{u}enza, L.F. \& Sig\"{u}enza, J.A.
{\it Phys. Rev. Lett.} {\bf 84}, 2758-2761 (2000).

\bibitem{Bohland01} Bohland, J.W.\& Minai, A.A.
{\it Neurocomputing} {\bf 38-40}, 489-496 (2001).

\bibitem{Labiouse02} Labiouse, C.L, Salah, A.A.\& Starikova,I., 
"The impact of connectivity on the memory capacity and the retrieval 
dynamics of Hopfield-type networks," in Proceedings of the 
Santa Fe Complex Systems Summer School, pp. 77-84, Budapest, NM: SFI, 2002.

\bibitem{Sola04}  Rosin, A., Riecke, H., \& Solla, S.A. 
{\it Phys. Rev. Letters} {\bf 92}, 19 (2004).

\bibitem{Netoff04} Netoff, T.I, Clewley, R., Arno, S., Keck, T. \& White, J.A.
{\it J. Neurosci} {\bf 24}, 8075-8083 (2004).

\bibitem{Sporns00} Sporns, O., Tononi, G., \& Edelman, G.M. 
{\it Cerebral Cortex} {\bf 10}, 127-141 (2000).

\bibitem{Dante04} Dante R. Chialvo, {\it Physica} {\bf A340}, 756-765 (2004).

\bibitem{Laughlin03} Laughlin, S.B. \& Sejnowski, T.J. {\it Science} {\bf 301},  1870-1874 (2003).

\bibitem{Rosenblatt62} Rosenblatt, F. {\it Principles of Neurodynamics}, 
Spartan, New York (1962).

\bibitem{LeCun98} LeCun, Y., Bottou, L., Bengio, Y. \& Haffner, P. 
{\it Proc. of the IEEE} {\bf 86}, 2278-2324 (1998). 

\bibitem{Rummelhart86} Rummelhart, D.E., Hinton, G.E. \& Williams,
R.J. {\it Nature} {\bf 323}, 533-536 (1986).

\bibitem{Latora} V.~Latora, M.~Marchiori {\it Phys. Rev. Lett.} {\bf 87}, 198701 (2001).

\bibitem{Marchiori00} Marchiori, M. \& Latora, V.
{\it Physica} {\bf A285}, 539-546 (2000).

%\bibitem{Hertz} Hertz, J., Krogh, A. \& Palmer, R.G.
%{\it Introduction to the Theory of Neural Computation}, Santa Fe
%Institute studies in the sciences of complexity, Lecture Notes Vol.
%I, Addison-Wesley Pub., Redwood City, California (1991).

\end{thebibliography}
\end{document}